\documentclass[prb,twocolumn,showpacs,preprintnumbers,amsmath,amssymb]{revtex4}
\usepackage{amsmath,amssymb,amsfonts}
\usepackage{bm}
\usepackage{graphicx}
\usepackage{times}
\usepackage{color}
\usepackage[colorlinks,bookmarks=false,citecolor=blue,linkcolor=red,urlcolor=blue]{hyperref}

\newcommand{\fref}[1]{Fig.~\ref{#1}}

\newcommand{\normwidth}{0.8\columnwidth}

\newcommand{\midwidth}{0.75\columnwidth}

\begin{document}

\title{Surface effects in doping a Mott insulator}
\author{Reza Nourafkan}
\affiliation{Department of Physics, University of Alberta, Edmonton, Alberta, Canada T6G 2G7}
\author{Frank Marsiglio}
\affiliation{Department of Physics, University of Alberta, Edmonton, Alberta, Canada T6G 2G7}

\begin{abstract} 
The physics of doping a Mott insulator is investigated in the presence of a solid-vacuum interface. Using the embedding approach for dynamical mean field theory we show that the change in surface spectral evolution in a doped Mott insulator is driven by a combination of charge transfer effects and enhanced correlation effects. Approaching a Mott insulating phase from the metallic side, we show that a dead layer forms at the surface of the solid, where quasiparticle amplitudes are exponentially suppressed. Surface correlation and charge transfer effects can be strongly impacted by changes of the hopping integrals at the surface.
\end{abstract}
\maketitle

\section{Introduction}

Recently, interest in the electronic properties of surfaces and interfaces of strongly correlated electron systems has grown due to the discovery of conducting interfaces, with a substantial high carrier mobility, between two insulating perovskites.\cite{Ohtomo02} Theoretical studies of these systems\cite{Okamoto04_1, Okamoto04_2, Okamoto04_3} concluded that the leakage of charge from one layer to the other explains the results. Thus, charge transfer between materials in heterostructures can be used to stabilize interface states that otherwise would be obtained only via the chemical doping of the parent oxide compound. On the other hand, comparing results from measurements with different surface-sensitive photoemission, it has been shown that the spectral function can be strongly modified at the surface, compared with the bulk.\cite{Maiti98, Maiti01, Maiti04} For example, it was shown that the weight of the coherent part of the spectral function is reduced at the surface.\cite{Maiti98, Maiti04} 

Interfaces obviously break translational invariance,  and where present, the rotational symmetry of the bulk. Beyond this significant alteration in symmetry, there are important quantitative changes, such as a reduction of dispersion and changes of the crystal fields. Stress or strain is induced by the interface, which alters the distances and bonds between the ions. The lattice may even be structurally or electronically reconstructed. Due to these and other phenomena, the interface shifts and distorts the electronic states and energy levels at the ions, and modifies the entire electronic band. Since the electronic states of the individual ions and the screening are altered at interfaces, their correlations are strongly modified as well.\cite{Mannhart05} These phenomena raise an important question: how does the surface/interface electronic phase differ from the bulk?

Theoretical research on the effects of a surface or interface on strongly correlated electrons in inhomogeneous systems was initiated in large part by Potthoff and Nolting,\cite{Potthoff99_1} who studied the Hubbard model using an extension of dynamical mean field theory (DMFT)\cite{Georges96} appropriate to a layered system with a surface. Studies of solid-vacuum interfaces have unveiled the possibility of surface ferromagnetism,\cite{Potthoff99_2} and explored the penetration depth of a bulk metallic phase into an otherwise insulating surface.\cite{Helmes08, Ishida09}  Borghi \textit{et al.} have indeed shown the existence of a dead layer, due to an exponential penetration of metallic excitations.\cite{Borghi09} Yunoki \textit{et al.}\cite{Yunoki07} have studied the possibility of magnetic ordering and superconductivity at the interface between undoped high-$T_{c}$ cuprates and manganites. Ishida \textit{et al.} have studied the effect of interplanar Coulomb correlations on the electronic structure of strongly correlated heterostructures by applying the cluster DMFT method.\cite{Ishida10}  Studies of solid-vacuum interfaces of the Holstein-like models in an inhomogeneous system have shown that the polaron crossover takes place first on the surface. \cite{Nourafkan09, Nourafkan10}

Most of this previous work has been done for the half-filled Hubbard model with particle-hole symmetry. However, at interfaces there is not only a change of the energy spectrum and of the correlation, but due to charge transfer, the state occupancy also varies. In the present work, we focus on charge transfer phenomena in the framework of a Hubbard-like model Hamitonian. Especially, we are interested in a doping-driven Mott transition in the presence of a surface.

The outline of this paper is as follows. In section II we introduce the model Hamiltonian, which is a semi-infinite Hubbard model with layer dependent parameters. The corresponding results are presented and analyzed in section III. Finally, section IV is devoted to concluding remarks.

\section{The Model and Method}
The simplest model Hamiltonian exhibiting Mott physics\cite{Mott49} is the Hubbard model,\cite{Hubbard63} in which the local repulsion correlates the electronic motion. At half-filling, this model captures many properties of the pressure-driven Mott transition. The Hubbard model away from half-filling can be used to investigate the density-driven Mott transition.
Here we try to model some part of the phenomena occurring at the interfaces of the strongly correlated electron systems using the Hubbard model with non-uniform model parameters. The system is a three-dimensional,
bipartite simple-cubic (sc) lattice with nearest-neighbor hopping only. The lattice is cut along a plane perpendicular to one
of the coordinate axes, e.g. the z-axis [sc(001) surface].
The system is considered to be built up by two-dimensional layers parallel to the surface.
Accordingly, the position vector to a particular site in the
semi-infinite lattice is written as ${\bm R}_{site}={\bm r}_{i}+{\bm
R}_{\alpha}$. Here ${\bm R}_{\alpha}$ stands for the coordinate
origin in the layer $\alpha$, and the layer index runs from
$\alpha=1$ for the topmost surface layer to infinity. The variable ${\bm r}_{i}$ is
the position vector with respect to a layer-dependent origin, and
runs over the sites within the layer. Each lattice site is then labeled by
indices $i$ and $\alpha$.
In this notation, the Hamiltonian reads:
\begin{eqnarray}
\label{eq:Holstein}
H=-\sum_{\langle i\alpha , j\beta\rangle \sigma} t_{i\alpha,j\beta} c^\dagger_{i\alpha \sigma}c_{j\beta \sigma} + \sum_{i\alpha}U_{\alpha} n_{i\alpha\uparrow}n_{i\alpha\downarrow},
\end{eqnarray}
where $c_{i\alpha\sigma}\left(c^\dagger_{i\alpha\sigma} \right)$ is the
destruction (creation) operator for electrons with spin
$\sigma$ on site $i$ of the $\alpha$ layer. The electron density on site $i\alpha$ is denoted $n_{i\alpha}$,  and $t_{i\alpha,j\beta}$ is the hopping
matrix element between two nearest-neighbor sites. $U_{\alpha}$ is the layer-dependent electron-electron on-site interaction. We use a uniform $U$ and fix the energy scale by setting $t_{\langle
i\alpha,j\beta\rangle}\equiv t = 1$ for $\alpha,\beta \neq 1$. The intralayer hopping $t_{i1,j1}=t_{11}$ or the hopping between the surface and the subsurface layer $t_{i1,j2}=t_{12}$ can be modified ($t_{11},t_{12} \ne t$) to simulate different correlation strengths and/or relaxation processes at the surface.

Our calculations are based on the embedding approach\cite{Ishida09,Nourafkan10} for dynamical mean field theory for a simple cubic lattice. This method is based on partitioning of the layered structure into a surface region which includes the first $N$ layers and an adjacent semi-infinite bulk region (substrate) to which it is coupled. Then the effect of the substrate on the surface region is described by an energy-dependent embedding potential. In constructing the embedding potential the self-energy of a bulk crystal corresponding to the substrate is required; this is obtained through a standard DMFT calculation. After constructing the embedding potential the self energy matrix of the surface region is determined using a layered DMFT calculation. In the framework of single-site DMFT, the self-energy matrix of the surface region is local [i.e., $\Sigma_{\alpha \beta}(i\omega_n)=\Sigma_{\alpha}(i\omega_n)\delta_{\alpha \beta}$] and independent of wave vector. Using the converged self-energy matrix we can find quasiparticle weights and local occupancies. 

In this study, the number of surface layers is chosen to be $N=5$ and we tested that this number provides converged results. Our impurity solver is exact diagonalization,\cite{Caffarel94} where the bath is represented in terms of a finite number of levels, $n_s$. The typical value we considered for the bath level is $n_s=9$ and we tested that this number provides essentially converged results. All errors associated with the use of these numbers are smaller than the points on the graphs.
 
\section{Results}

For a bulk lattice, in a phase with translational invariance, local occupations are site independent, $\langle n_{\alpha\uparrow}\rangle +\langle n_{\alpha\downarrow}\rangle = \langle n_{\alpha}\rangle=n$.
On the contrary, for a semi-infinite lattice, the different local environment of the surface sites causes the local occupations near the surface to differ from the bulk filling. Charge transfer is present in the free-electron system\cite{Kienert07} and we discuss the facts already known for this case first. The reduced coordination number and probably different 
hopping elements connecting at least one site in the surface layer of the semi-infinite system leads to a narrowing of the surface local density of states (LDOS), $\rho_{\alpha=1}^{(0)}(E)$, relative to the bulk, $\rho_{bulk}^{(0)}(E)$ (spectral weight is reduced at low and high energies but enhanced at intermediate energies). However,  the surface LDOS and the bulk LDOS have the same band edges and their overall width is given by the free bulk DOS.\cite{Kalkstein71} The band narrowing can be understood by referring to the moments of the LDOS, 
\begin{equation}
M_{i\alpha}^{(m,0)}=\int_{-\infty}^{\infty} E^m \rho_{i\alpha}^{(0)}(E)dE.
\label{moments}
\end{equation}
For a site in the bulk, the width of the DOS is given by  $\Delta \rho_{bulk}^{(0)}=M^{(2,0)}-(M^{(1,0)})^2=6t^2$, while for a site on the surface the result is  $\Delta \rho_{\alpha=1}^{(0)}=4t_{11}^2+t_{12}^2$.\cite{Potthoff97} Clearly, if these hopping matrix elements are equal to those in the bulk, then this width is narrower at the surface than at the bulk. From this fact, it follows that in order to ensure thermodynamic equilibrium, i.e., a common chemical potential $\mu$, one has $n_{\alpha=1}<n$ below and $n_{\alpha=1}>n$ above half-filling, since  the surface electron density  $2\int_{\infty}^{\mu} \rho_{\alpha=1}^{(0)}(E)dE$ is lesser (greater) for below (above) half filling than the bulk density, $2\int_{\infty}^{\mu} \rho_{bulk}^{(0)}(E)dE$. Thus the surface layer will have charge depletion (accumulation) below (above) half filling.\cite{Kienert07} 

These effects are all due to a reduced coordination number and probably different 
hopping elements. In addition a strong electron-electron interaction can cause a quite complicated charge redistribution at the surface. In the following, we present results on charge transfer and its effects on spectral weight for a semi-infinite Hubbard model with uniform- and non-uniform hopping elements.

\subsection{Uniform parameters}
\begin{figure}
\begin{center}
\center{\includegraphics[width=\midwidth]{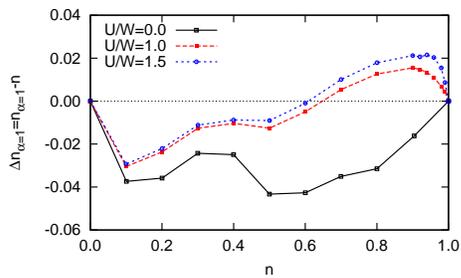}}
\caption{(Color online) The charge transfer $\Delta n_{\alpha=1}= n_{\alpha=1}-n$ between the surface layer ($\alpha=1$) and the bulk as a function of the filling $n$ for uniform $t$.}\label{fig:ChargeTransferUniform}
\end{center}
\end{figure}

We start our discussion with the case of uniform parameters. \fref{fig:ChargeTransferUniform} shows the charge transfer $\Delta n_{\alpha=1}= n_{\alpha=1}-n$ between the surface layer ($\alpha=1$) and the bulk as a function of the filling $n$ at $T=0$. The three curves are for (a) $U=0$ as just discussed, (b) $U/W = 1$, a value chosen to be below the critical Mott transition value $U_c/W \approx 1.33$,\cite{Borghi09} and (c) $U/W = 1.5$, a value chosen to be above the critical Mott transition value $U_c$. 

As argued above, the case without interactions displays surface charge depletion for all electron densities below half filling (and exactly the opposite for above half filling). For cases where there are interactions between the electrons, the situation is clearest in the strong coupling limit, where, at least at half filling, there are two well separated Hubbard bands. Insofar as doping into the $n< 1$ regime involves only the lower Hubbard band (definitely an oversimplification) then we should see an electron density dependence over this entire electron density regime ($0< n < 1$) similar to what is observed in the case without interactions, but over the entire density range ($0<n<2$). This is borne out by the results in \fref{fig:ChargeTransferUniform} for non-zero interaction strength: a surface electron density depletion is observed for small to intermediate densities, and an accumulation for densities from beyond quarter filling to near half filling is seen. Apparently for an interacting system the key effect of a narrowing of the surface local density of states holds quite generally, even in the presence of interactions, and even within the lower and upper Hubbard bands. 

Upon increasing the bulk charge density, the chemical potential moves from the lower half of the lower quasiparticle sub-band to the upper half of that same sub-band. In addition, the differences between the $n_{\alpha}$ and the bulk $n$ diminish with increasing distance from the surface and for  the third layer, the occupation is almost indistinguishable from the bulk occupation (not shown). Finally, note that the charge transfer in the noninteracting limit is larger than in the interacting cases, so the expectation that increased interactions tend to reduce overall the magnitude of charge transfer is supported by these results. Indeed, a finite charge transfer implies an (unfavorable) increase of the average double occupancy compared with the homogeneous charge distribution; thus, a non-zero Hubbard interaction between conduction electrons will tend to suppress charge transfer. However, due to separation of low- and high-energy states, charge transfer is not completely suppressed in the strong coupling limit, $U \rightarrow \infty$, even when the long range Coulomb interaction is present.   

Also as expected, at half-filling there is no charge transfer at all, because particle-hole symmetry on a bipartite lattice with nearest-neighbor hopping requires a homogeneous charge distribution among the layers.\cite{charge_transfer}

Theoretical studies of the Hubbard model for bulk systems, mainly using DMFT,\cite{Georges96} have shown that in a bulk system at half-filling, 
the system is insulating for a sufficiently large Coulomb repulsion, $U>U_{c}$. For any other filling, the ground state is metallic. In the metal the charge compressibility $\kappa=(1/n^2)(\partial n/ \partial \mu)$ remains finite even when the insulator is approached from above or below half-filling, but the chemical potential changes discontinuously from $\mu=\mu^-_{c}$ to $\mu=\mu^+_{c}$ as the filling goes from $n=1^-$ to $n=1^+$. For $\mu^-_{c}<\mu<\mu^+_{c}$, the system is insulating and half-filled. Doping away from half-filling induces states inside the Mott-Hubbard gap.\cite{Fisher95,Kajueter96} Therefore the jump in chemical potential $\Delta \mu = \mu^+_{c} - \mu^-_{c}$ for infinitesimal doping is strictly less than the single-particle gap of the insulator.

Another feature of the DMFT treatment of the Mott transition is that both the interaction-driven transition at half-filling\cite{Georges96} and the density-driven transition\cite{Ono02, Eckstein07, Werner07, Liebsch08, Sordi10} are characterized by the coexistence of two solutions in some parameter regimes. In the latter case, there is a region of chemical potentials, delimited by two curves $\mu^-_{c1}(U)$ and $\mu^-_{c2}(U)$ for hole doping, in which a metallic solution with $n \neq 1$ and an insulating one with $n=1$ coexist. Naturally, the stable solution is the one that minimizes the grand-canonical free energy. At $T=0$ the metal is stable in the whole coexistence region and is continuously connected to the insulator. At finite temperature the energetic balance is more complicated, and a first-order transition occurs along an intermediate line between $\mu_{c1}$ and $\mu_{c2}$, leading to phase separation between the two phases (formation of a phase mixture with a fraction $x_{met}$ of the metal and $1-x_{met}$ of the insulator). 
 
\begin{figure*}
\begin{center}
\center{\includegraphics[width=1.6\columnwidth]{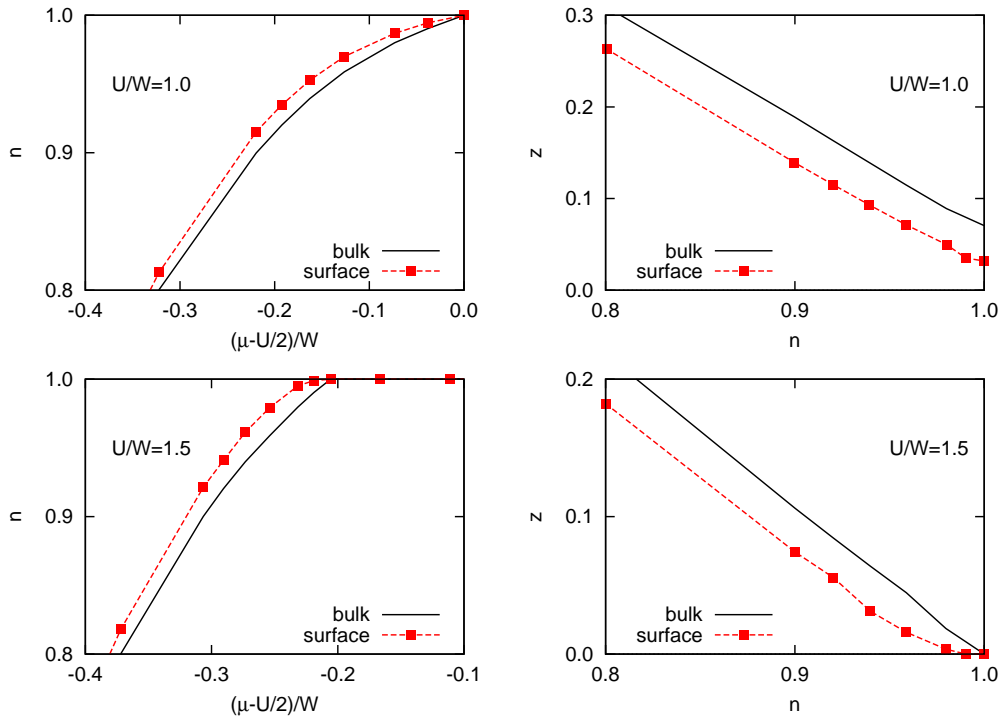}}
\caption{(Color online) Left panels: Occupation $n$ vs $(\mu-U/2)/W$ for two different values of interaction strength $U/W$. The data shown are for the bulk system and the surface layer with uniform $t$. The plateau in the occupation at $n=1$ signals the onset of the incompressible Mott state. Right panels: Surface layer quasiparticle weight $z_1$ for uniform $t$ and the bulk quasiparticle weight as a function of occupation. The solid line shows $z$ for the bulk calculation.}\label{fig:DensityUniform}
\end{center}
\end{figure*}

Is it possible to drive a Mott transition on the surface at a critical chemical potential different from that required for the bulk? To address this question, we focus on the behaviour of the electron density $n$ as a function of the chemical potential $\mu$ and on the density-driven Mott transition. A plot of the occupation as a function of the chemical potential, shown in the left panels of \fref{fig:DensityUniform} for two different values of $U$, clearly shows the Mott transition into an incompressible state, as a plateau near $\mu = U/2$, when the Hubbard $U$ is greater than the critical coupling $U_c$. When $U$ is lower than the critical value (upper left panel) then, while the filling still approaches unity at $\mu = U/2$, there is clearly no plateau present. 

As can be seen in the lower left panel, for chemical potential near the critical value (where $n$ becomes unity), the surface layer approaches an insulating state slightly faster than the bulk. Nonetheless, the electron density eventually approaches unity at the {\em same} chemical potential that governs the bulk density. The net result is that the surface tracks the bulk as far as the metal-insulator transition is concerned, and a metal in the bulk implies a metal at the surface.    

To examine this transition in more detail, the right panels of \fref{fig:DensityUniform} show the quasiparticle weight for the $\alpha^{th}$ layer, $z_{\alpha}=\left( 1-\partial\Sigma_{\alpha}(\omega)/\partial\omega\vert_{\omega=0}\right)^{-1}$ ($\Sigma_{\alpha}(\omega)$ is the self-energy for layer $\alpha$) as a function of occupation for $U/W=1.0$ and  $U/W=1.5$. This figure shows the existence of a so called dead layer at the surface characterized by an exponentially decaying quasi-particle weight near half-filling (lower panel). In the upper panel, the quasiparticle weight in the bulk does not reach zero, indicating that the $U$ is below the critical value. Smaller $z$ at the surface at small doping means the coherent peak near the Fermi energy is narrower at the surface and the incoherent Hubbard band is more pronounced than in the bulk, in agreement with experiment.\cite{Maiti98, Maiti04} Such a trend has been reported in studies of the correlation-driven metal-insulator phase transition at half-filling.\cite{Potthoff99_2} However, the reduction of the surface quasiparticle weight seen here is much weaker than in the experiments.   

At low electron density the quasiparticle weight for the surface layer is greater than for the bulk (not shown). In this region the depletion of charge from the surface to the bulk (see \fref{fig:ChargeTransferUniform}) causes a reduction of correlation effects, which leads to a more robust quasiparticle at the surface. 

\begin{figure}
\begin{center}
\center{\includegraphics[width=\normwidth]{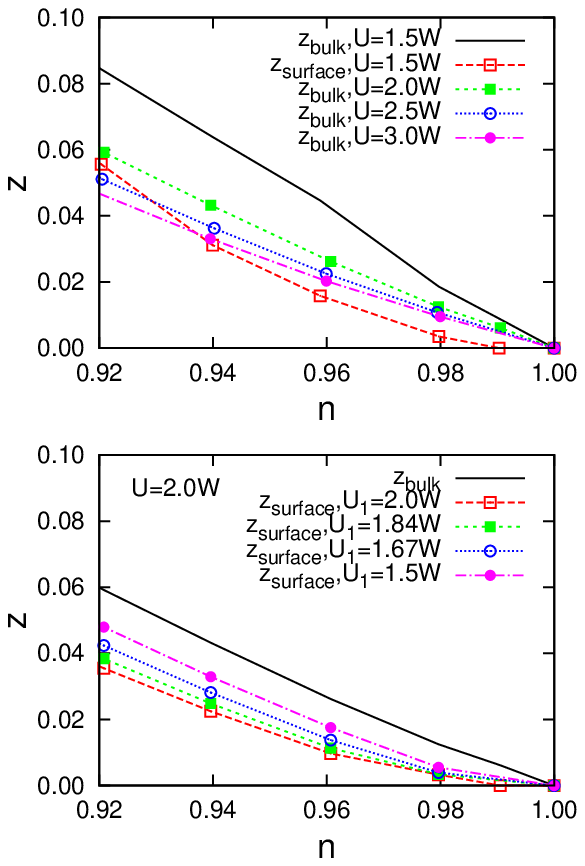}}
\caption{(Color online) Plot of the quasiparticle weight, $z$ vs. electron density, $n$, near half filling for $U/W = 1.5$ (top panel), and $U/W = 2.0$ (bottom panel). The bulk result is given by the solid (black) curve. The corresponding result for the surface layer is given by the red (dashed) curve with open square symbols; the value of $z$ at the surface is significantly reduced from that in the bulk. Since the coordination number is lower on the surface then the surface electrons effectively experience a large Coulomb repulsion. If this is the primary reason for the reduced value of $z$, then a bulk calculation with a larger value of $U$ should mimic the original result at the surface. This is clearly not the case, even when we double the value of $U$. In the right panel we instead lower the value of $U_1$ (on the surface) in an attempt to restore a value of $z$ equal to that in the bulk; this is also unsuccessful, indicating that charge transfer effects are an important ingredient for the
reduced quasiparticle weight on the surface.}
\label{fig:QuasiParticleUniform}
\end{center}
\end{figure}

To illustrate to what extent charge transfer is responsible for the differences at the surface vs. the bulk, we try to eliminate the effect of enhanced correlation at the surface. In the left panel of \fref{fig:QuasiParticleUniform} we show the bulk and surface quasi particle weights for $U/W=1.5$ as a function of electron density very close to half-filling. This figure also include several bulk calculations for larger values of $U$. If the enhanced correlation effect is responsible for the differences at the surface, we should recover the original value of the surface $z$ in a bulk calculation with a larger value of $U$. As is seen from the figure, even for a very large value of $U$, the bulk calculation is unable to reproduce the original surface result. Therefore charge transfer effects play an important role in a doped Mott insulator. As another test for this statement, in the right panel of
\fref{fig:QuasiParticleUniform} we use $U/W = 2.0$ in the bulk, and then show several semi-infinite calculations with reduced surface correlation $U_1$. This figure also shows that upon reducing the value of $U_1$ it is not possible to have the same quasiparticle weights at the surface as in the bulk calculation. Thus, the charge transfer effect is more important than enhanced correlation effects in a doped Mott insulator. This is expected because the interaction strength is already large enough to give a Mott insulator at half-filling, so enhancing that value would not lead to any significant change. Indeed, the occupation distance from half-filling is more important here. This quantity tunes with the charge transfer effect.

\subsection{Modified surface parameters}
\begin{figure}
\begin{center}
\center{\includegraphics[width=\midwidth]{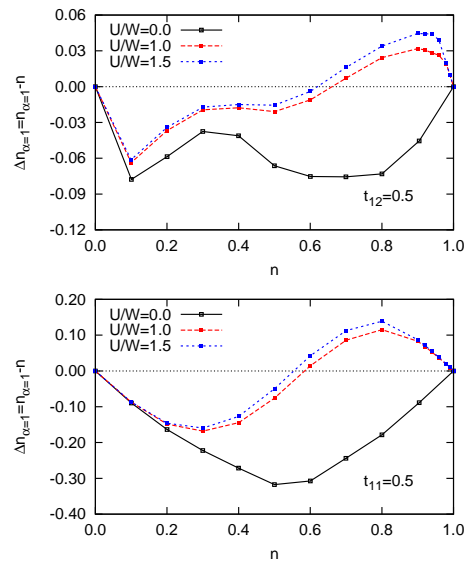}}
\caption{(Color online) The charge transfer $\Delta n_{\alpha=1}= n_{\alpha=1}-n$ between the surface layer ($\alpha=1$) and the bulk as a function of the filling $n$ for $t_{12}/t=0.5$ and $t_{11}/t=0.5$ at $T=0.0$.}
\label{fig:ChargeTransferNonUniform}
\end{center}
\end{figure}

In the previous section, we mentioned that the reduction of the quasiparticle weight at the surface is smaller when compared to the reduction found in experiments. Here we show that a modified hopping at the surface causes a larger charge transfer and then the reduction in surface quasiparticle weight is greatly amplified. In 
\fref{fig:ChargeTransferNonUniform} we show the charge transfer for the case where the hopping from a site on the surface to one on the layer beneath is lower, $t_{12}=0.5$, and the case where the intralayer hopping on the surface layer is reduced with respect to that in the bulk, $t_{11}=0.5$. These figures show that reduced hopping matrix elements on the surface sites enhance the amount of charge transfer. In the case of $t_{11}=0.5$, the value of charge transfer is sufficiently large that including the long-range Coulomb interaction will have some effect in reducing the total amount. For simplicity, however, we exclude this point here; we believe that including the long-range Coulomb interaction will lessen some of the effects described below, but the basic results will still hold. 

\begin{figure*}
\begin{center}
\center{\includegraphics[width=1.6\columnwidth]{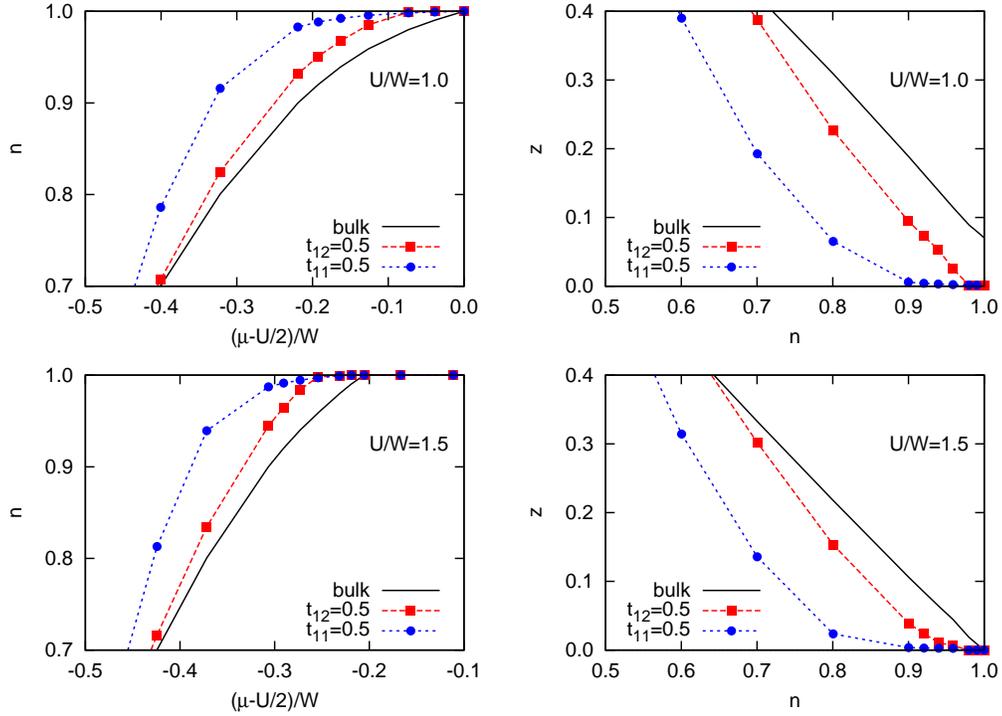}}
\caption{(Color online) Left panels: Occupation $n$ vs $(\mu-U/2)/W$ for two different values of interaction strength $U/W$. The data shown are for the bulk system and the surface layer with $t_{12}/t=0.5$ and $t_{11}/t=0.5$. The plateau in the occupation at $n=1$ signals the onset of the incompressible Mott state. Right panels: Surface layer quasiparticle weight $z_1$ for non-uniform $t$ and the bulk quasiparticle weight as a function of occupation. The solid line shows $z$ for the bulk calculation.}\label{fig:DensityNonUniform}
\end{center}
\end{figure*}

\fref{fig:DensityNonUniform} shows the electron density as a function of the chemical potential (left panels) and quasiparticle weights  as a function of electron density (right panels) for reduced hopping parameters near the surface. As can be seen in \fref{fig:DensityNonUniform} the tendency toward an insulating state is certainly enhanced as we lower the hopping parameters near the surface. For reduced parameters, particularly for reduced $t_{11}$, the plateau region is considerably extended (left panels). Close examination, however, verifies that even in the case of non-uniform parameters, there is only one transition for the entire sample. Also, results for quasiparticle weights (right panels) show that the reduction in $z$ is considerably more pronounced in the case of reduced parameters and is almost negligible for a wider range of doping.  It is a matter for experiment to decide if such
changes in surface parameters can explain the data in a consistent manner.

\begin{figure}
\begin{center}
\center{\includegraphics[width=\normwidth]{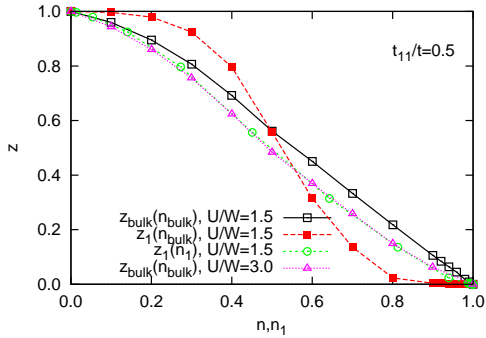}}
\caption{(Color online) The quasiparticle weight is plotted vs electron density. The open squares joined by solid (black) lines denote the bulk weight as a function of the bulk electron density; this shows a characteristic decrease from $n \approx 0$ to half-filling, where an insulating state is present. The surface quasiparticle weight, $z_1$ (filled red squares joined by dashed (red) lines) is shown vs. the same bulk electron density, and shows considerable variation from the bulk. The same quantity, but plotted vs. the surface electron density, is shown by open circles joined by (green) short-dashed lines.  This result is in qualitative agreement with the bulk result. An improvement is achieved by accounting for the effective higher Coulomb repulsion, as described in the text; this is shown with open triangles joined by (pink) dot-dashed lines; this result is in excellent quantitative agreement with the renormalized surface quasiparticle weights. We have used $t_{11}/t = 0.5$ for the calculations shown in this plot.} 
\label{fig:QuasiParticle2}
\end{center}
\end{figure}

To further illustrate to what extent charge transfer is responsible for the differences at the surface vs. the bulk, even in cases where the surface parameters can vary from those in the bulk, we reshow in 
\fref{fig:QuasiParticle2} the results for $z$ for $U/W = 1.5$ and $t_{11}/t = 0.5$ for both the bulk (open squares joined by solid (black) lines) and the surface (filled red squares joined by dashed (red) lines). There are clear and significant differences, particularly at low ($n \approx 0$) and high ($n \approx 1$) electron concentration; however, to account for charge transfer effects we use the information in \fref{fig:ChargeTransferNonUniform} and show the surface quasiparticle weight on the same plot as a function of $n_1$; this is shown by open circles joined by (green) short-dashed lines. The trend as a function of electron concentration now is qualitatively similar to that of the bulk. However, one can improve upon this even further, since, in reality the electrons on the surface experience a stronger relative Coulomb repulsion due to the reduced intralayer hopping parameter. For this reason we include a calculation in the bulk for $U/W = 3.0$ (since the intralayer hopping in this layer is half what it is elsewhere); this is shown with open triangles joined by (pink) dot-dashed lines, and these results are in excellent quantitative agreement with the renormalized surface quasiparticle weights.\cite{remark2} Thus, we find that essentially the entire change of the quasiparticle weight at the surface can be accounted for once the charge transfer effects are taken into consideration.

\section{Concluding Remarks}

In conclusion we have investigated the surface electronic properties of a doped Mott insulator by applying the embedding approach for DMFT. As a model system, we considered a semi-infinite Hubbard model. We have shown that the reduced coordination number and probably reduced hopping elements  at the surface causes an accumulation of charge at the surface in the strong coupling limit, that makes the surface approach the condition of half-filling more quickly than the bulk. The accumulation of charge at the surface, in turn, makes the correlation effects more effective. In particular, we have demonstrated that the reduction of the quasiparticle weight detected by surface sensitive photoemission experiments of a doped Mott insulator are caused by both charge transfer and enhanced correlation effects at the surface. The expected modification of the intra-layer hopping at the surface and inter-layer hopping between the surface and the subsurface layer lead to more significant differences between the surface and the bulk.
 
\begin{acknowledgements}
This work was supported in part by the Natural Sciences and Engineering
Research Council of Canada (NSERC), by ICORE (Alberta), and by the Canadian
Institute for Advanced Research (CIfAR).

\end{acknowledgements}

\bibliographystyle{prsty}

\begin{thebibliography}{10}
\bibitem{charge_transfer}
For a non-bipartite (e.g. fcc) lattice a charge modulation is possible also at half-filling.
\bibitem{strong_coupling}
In the strong-coupling limit the charge density is almost homogeneous in the inner layers of the system and only the surface charge density $n_{\alpha=1}$ is enhanced significantly.
\end{thebibliography}

\end{document}